\documentclass[aps,prl,twocolumn,superscriptaddress,epsfig,amsmath,showpacs]{revtex4}
\usepackage{epsfig}
\usepackage{dcolumn}
\usepackage{bm}
\begin{document}
\title{Fermi surfaces and quantum oscillations in underdoped high-$T_c$ 
superconductors YBa$_2$Cu$_3$O$_{6.5}$ and YBa$_2$Cu$_4$O$_8$}

\author{Hyungju Oh}
\affiliation{Department of Physics and IPAP, Yonsei University, 
Seoul 120-749, Korea}
\author{Hyoung Joon Choi}
\email[Email:\ ]{h.j.choi@yonsei.ac.kr}
\affiliation{Department of Physics and IPAP, Yonsei University, 
Seoul 120-749, Korea}
\author{Steven G. Louie}
\affiliation{Department of Physics, University of California, 
Berkeley, California 94720, USA}
\affiliation{Materials Sciences Division, Lawrence Berkeley 
National Laboratory, Berkeley,California 94720, USA}
\author{Marvin L. Cohen}
\affiliation{Department of Physics, University of California, 
Berkeley, California 94720, USA}
\affiliation{Materials Sciences Division, Lawrence Berkeley 
National Laboratory, Berkeley,California 94720, USA}

\date{\today}

\begin{abstract}
We study underdoped high-$T_c$ superconductors YBa$_2$Cu$_3$O$_{6.5}$ and 
YBa$_2$Cu$_4$O$_8$ using first-principles pseudopotential methods
with additional Coulomb interactions at the Cu atoms, and obtain
Fermi-surface pocket areas in close agreement with measured Shubnikov-de 
Haas and de Haas-van Alphen oscillations. With antiferromagnetic order 
in CuO$_2$ planes, stable in the calculations, small hole pockets are 
formed near the so-called Fermi-arc positions in the Brillouin zone which 
reproduce the low-frequency oscillations. A large electron pocket, 
necessary for the negative Hall coefficient, is also formed in 
YBa$_2$Cu$_3$O$_{6.5}$, giving rise to the high-frequency oscillations 
as well. Effective masses and specific heats are also calculated and 
compared with measurements. Our results highlight the important role of 
magnetic order in the electronic structure of underdoped high-$T_c$ 
superconductors.
\end{abstract}


\maketitle

The normal-state electronic structures of the underdoped high-$T_c$ 
superconductors have been studied for more than twenty years, 
but the Fermi-surface (FS) topology is still only partially 
understood\cite{norman,hossain,doiron,jaudet,sebastian,audouard,bangura,yelland,carrington,elfimov,puggioni,leboeuf,harrison,morinari,chen,chakravarty,podolsky}.
An important observation is the disconnected FSs\cite{norman,hossain}, 
namely Fermi arcs, observed in angle-resolved photoemission 
spectroscopy (ARPES), which initiated intense investigations about 
whether the FSs are really disconnected arcs or closed pockets 
of which one side is hardly visible in ARPES. Recently, in contrast 
to having Fermi arcs, de Haas-van Alphen (dHvA) oscillations in the 
magnetization and Shubnikov-de Haas (SdH) oscillations in the 
resistance\cite{doiron,jaudet,sebastian,audouard,bangura,yelland}
observed in ortho-II YBa$_2$Cu$_3$O$_{6.5}$ and YBa$_2$Cu$_4$O$_8$ suggest 
well-defined close pockets in the FS of underdoped high-$T_c$ cuprates. 
The measured oscillations for YBa$_2$Cu$_3$O$_{6.5}$ are a dominant one 
at 500$\pm$20~T with a satellite at 1650$\pm$40~T \cite{sebastian}, 
and more recently a dominant oscillation at 540$\pm$15~T with 
satellites at 450$\pm$15~T, 630$\pm$40~T, 
and 1130$\pm$20~T \cite{audouard}. For YBa$_2$Cu$_4$O$_8$, 
oscillation at 660$\pm$15~T is observed\cite{bangura,yelland}.

The measured dHvA and SdH oscillations provide extreme cross-sectional 
areas of closed FS pockets\cite{onsager}, but they alone are not enough 
to identify the shapes and locations of the pockets. Thus, a 
quantitative theoretical calculation of the FS geometry can be useful 
to determine the FS topology. First-principles calculations based on 
the density functional theory (DFT) approach have been performed for 
YBa$_2$Cu$_3$O$_{6.5}$ and 
YBa$_2$Cu$_4$O$_8$ \cite{carrington,elfimov,puggioni}, but the calculated 
FSs could not explain the oscillation measurements. In contrast to the 
meansurements, reported DFT calculations predict only FS pockets 
much larger than 500 T, and do not obtain the electron-type carriers 
implied by the observed negative Hall coefficients\cite{leboeuf}.

According to model 
calculations\cite{harrison,morinari,chen,chakravarty,podolsky}, 
antiferromagnetic (AFM) order or a $d$-density wave may result in 
small pockets in regions of the FS. Although using DFT calculations, 
one may examine static magnetic order using spin-density functional 
theory\cite{zhang}; as yet, no magnetic order has been considered in 
the reported DFT calculations of YBa$_2$Cu$_3$O$_{6.5}$ and 
YBa$_2$Cu$_4$O$_8$.

In this Letter, we present, for the first time, first-principles 
spin-density functional calculations of YBa$_2$Cu$_3$O$_{6.5}$ and 
YBa$_2$Cu$_4$O$_8$ with a Coulomb repulsion $U$ at Cu sites which 
yield FSs consistent with the dHvA and SdH measurements. It is shown 
that, with physically reasonable $U$ values, the AFM order in the 
CuO$_2$ planes reconstructs the FS and produces pockets with sizes 
consistent with the measured frequencies. Moreover, the calculated 
FS of YBa$_2$Cu$_3$O$_{6.5}$ has a large electron pocket which 
explains the observed negative Hall coefficients. In addition, 
cyclotron effective masses and specific heats are calculated and 
compared with experiments. Our results support the possible importance 
of magnetic order in the electronic structures of underdoped 
high-$T_c$ cuprates.

Our present work is based on {\em ab-initio} pseudopotential 
density-functional calculations with pseudo-atomic orbitals to expand 
the electronic wavefunctions\cite{sanchez}. Coulomb interaction at 
Cu $d$ orbitals, parameterized by $U$ and $J$ \cite{liechtenstein,park}, 
is added to the local (spin) density approximation [L(S)DA+$U$]. 
With experimental atomic structures\cite{grybos,yamada},
we minimize the total energy with respect to the magnetic moments of 
Cu atoms in the CuO$_2$ planes and CuO chains to consider the 
possibility of AFM order. Our results are that YBa$_2$Cu$_3$O$_6$ 
is an AFM insulator and YBa$_2$Cu$_3$O$_7$ is a non-magnetic metal.


\begin{figure} 
\centering
\epsfig{file=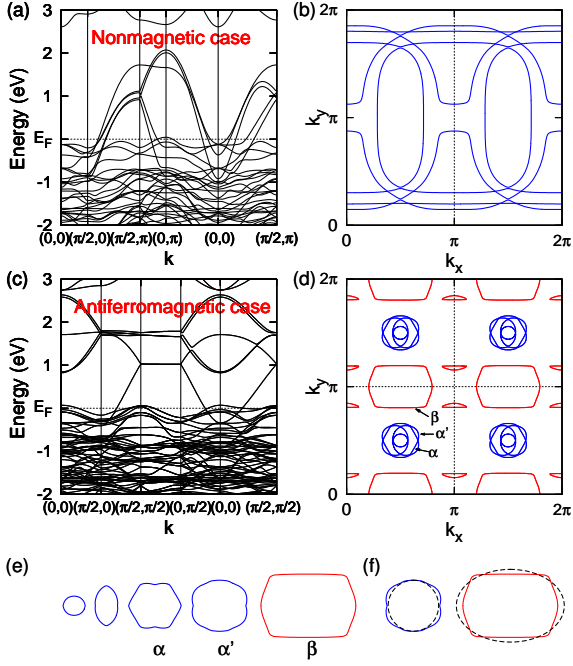,width=7.5cm,angle=0,clip=} 
\caption{(Color online)
Electronic structures in ortho-II YBa$_2$Cu$_3$O$_{6.5}$ with Coulomb 
interaction at Cu $d$ orbitals ($U$ = 6.0 eV and $J$ = 1.0 eV). 
(a) The band structure and (b) the FS obtained by the LDA+$U$ method 
with no magnetic order. (c) The band structure and (d) the FS obtained 
by the LSDA+$U$ method with a checker-board-type AFM order in 
the CuO$_2$ planes. In (b) and (d), the FSs are drawn in the Brillouin 
zone of a real-space unit cell (0.383 $\times$ 0.387 nm$^2$)
containing a Cu atom on each CuO$_2$ plane. Blue (red) lines are 
hole pockets and open orbits (electron pockets). (e) FS pockets in (d). 
The $\alpha$, $\alpha'$, and $\beta$ pocket areas are 485, 621, and 
1450~T, respectively. (f) Comparison with experimental FS pocket 
areas (630~T \cite{audouard} and 1650~T \cite{sebastian}) in dashed 
lines. A FS area of 1~nm$^{-2}$ corresponds to a frequency of 
105~T \cite{onsager}.
}
\end{figure}

Using the LDA+$U$ method with no magnetic order, we obtain the electronic 
structure for YBa$_2$Cu$_3$O$_{6.5}$ (Figs.~1a and 1b), which are in 
good agreement with previous calculations\cite{carrington,elfimov}. The 
FS has only large hole pockets and open orbits (Fig.~1b); however, 
this is not in agreement with the observed quantum oscillations.

When magnetic order is considered in the LSDA+$U$ calculations for 
YBa$_2$Cu$_3$O$_{6.5}$, a checker-board-type AFM order is stabilized in the 
CuO$_2$ planes, and this drastically changes the electronic structures 
(Figs.~1c and 1d). With $U$ = 6.0 eV and $J$ = 1.0 eV, each Cu atom in 
the CuO$_2$ plane has 0.48 Bohr magneton ($\mu_B$), while the CuO chains 
are still non-magnetic. The FS (Fig.~1d) now consists of small hole pockets 
($\alpha$ and $\alpha'$ indicating the two largest ones) and a large 
electron pocket ($\beta$). The calculated pocket 
areas, which are not very sensitive to $U$ and $J$ around the used 
values, are in good agreement with the experimental 
observations (Fig.~1f). This shows that the AFM order may be a way 
to quantitatively explain the measured dHvA and SdH frequencies.

Figure~1d shows that the hole pockets ($\alpha$ and $\alpha'$) are located 
at ($\pm \frac{\pi}{2}$, $\pm \frac{\pi}{2}$), close to the positions of 
the Fermi arcs in the ARPES data\cite{norman,hossain}. This supports the 
idea that slow AFM fluctuation\cite{hinkov_kampf_birgeneau_hayden_lake} 
may form hole pockets near ($\pm \frac{\pi}{2}$, $\pm \frac{\pi}{2}$), 
with their shapes possibly modified to form the arcs because of 
short-range fluctuation\cite{harrison,morinari}.

Figure~1d also shows that the electron pocket ($\beta$) is much more 
anisotropic than the hole pockets ($\alpha$ and $\alpha'$). This arises 
because the electron pocket is derived from the CuO chain and the 
CuO$_2$ plane, while the hole pockets come from the CuO$_2$ plane only.
To have an isotropic resistivity as observed in experiments at 
temperature below 80 K \cite{doiron,yando}, we find that the electron 
mean free path along the chain direction should be about one quarter 
of that perpendicular to the chain because of directional difference 
in the group velocity.


\begin{figure} 
\centering
\epsfig{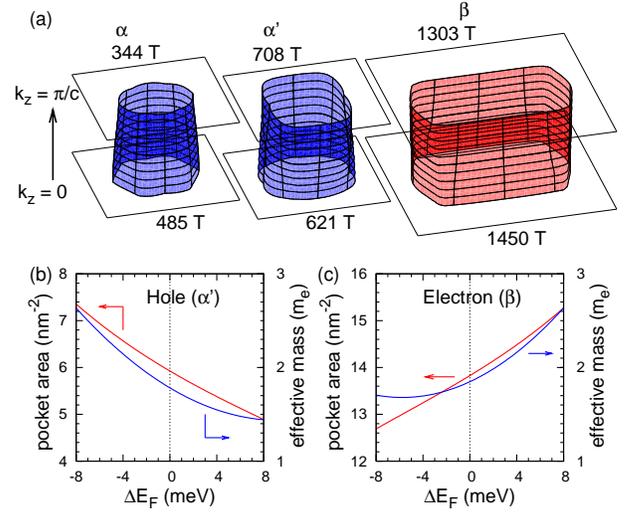} 
\caption{(Color online)
FS pocket sizes versus $k_z$ and cyclotron effective masses in 
YBa$_2$Cu$_3$O$_{6.5}$. (a) FS pocket areas perpendicular to the 
$k_z$-axis when the Fermi level is at the charge-neutrality level. 
(b) Hole and (c) electron pocket areas and their cyclotron effective 
masses as functions of the Fermi-level shift ($\Delta E_F$) from 
the charge-neutrality level. The indexes $\alpha$, $\alpha'$, and 
$\beta$ are the same as in Fig.~2d. A FS area of 1~nm$^{-2}$ 
corresponds to a frequency of 105~T \cite{onsager}.
}
\end{figure}

For more detailed comparison, Fig.~2a shows the $k_z$ dependence 
of FS pocket areas. The three largest extreme areas of hole pockets 
are 485 and 621~T at $k_z$=0 and 708~T at $k_z$=$\pi$/c (Fig.~2a), 
which  overestimate by about 15~\%  the experimental 
low frequencies, 450, 540, and 630~T \cite{audouard}.
The largest extreme area of the electron pocket ($\beta$) is 
1450~T at $k_z$=0 (Fig.~2a), and it underestimates by 12~\% the 
experimental high frequency, 1650~T \cite{sebastian}. If the Fermi 
level is shifted to higher energy, hole pockets would shrink and 
the electron pocket would expand (Figs.~2b and 2c). With a Fermi-level 
shift ($\Delta E_F$) of 4 meV above the charge-neutrality level, 
the extreme pocket sizes become 441, 564, and 652~T for the holes 
and 1519~T for the electron, respectively, resulting in closer 
agreement with experimental results.

We calculate the cyclotron effective masses (Figs.~2b and 2c) and the 
electronic contribution to the normal-state specific heat from the 
LSDA+$U$ electronic structure of YBa$_2$Cu$_3$O$_{6.5}$. The obtained 
cyclotron effective masses are 1.78 times the free electron 
mass ($m_e$) for the $\alpha'$ pocket and 1.88 $m_e$ for the $\beta$ 
pocket. These values are smaller than measured values, 
1.78 $\sim$ 1.9 $m_e$ for the low frequency and 3.8 $m_e$ for the 
high frequency\cite{doiron,jaudet,sebastian}, but they are consistent 
with experiments in the sense that the effective mass of the low-frequency 
oscillation (from the $\alpha'$ pocket in our result) is smaller than that 
of the high-frequency oscillation (from the $\beta$ pocket in our result). 
The calculated Sommerfeld coefficient for the normal-state specific heat is 
9.28 mJ mol$^{-1}$ K$^{-2}$ which slightly underestimates the 
experimental value of 10 mJ mol$^{-1}$ K$^{-2}$ \cite{loram}. The 
differences between our values and the measured ones may originate from 
many-body effects.

The presence of the electron pocket in our FS (Fig.~1d) definitely 
opens a chance of a negative Hall coefficient, but it alone 
is not sufficient since the total numbers of holes and electrons in 
our calculation are equal to each other to represent a charge-neutral 
stoichiometric sample. Since the Hall coefficient is inversely 
proportional to the net charge of the carriers, a slight imbalance 
of the two types of carriers would result in a relatively large Hall 
coefficient. With $\Delta E_F$ = 4 meV, as discussed above for the 
oscillation frequencies, we can obtain a negative Hall coefficient 
of -25 mm$^3$C$^{-1}$ at 70~T (Fig.~3a), close to the experimental 
value of about -30 mm$^3$C$^{-1}$ \cite{leboeuf}.


\begin{figure} 
\centering
\epsfig{file=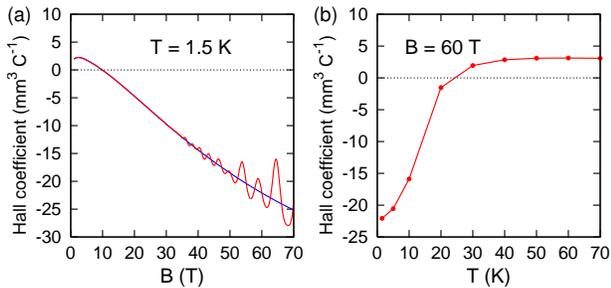,width=8cm,angle=0.0,clip=}  
\caption{(Color online)
Hall coefficients for YBa$_2$Cu$_3$O$_{6.5}$. (a) Hall coefficient 
versus magnetic field (B) at temperature $T$ =  1.5 K, with and 
without the SdH oscillation (red and blue lines, respectively). 
(b) Hall coefficient versus temperature at B =  60 T without 
considering the SdH oscillation.
}
\end{figure}

Figure 3 shows the Hall coefficients obtained by semiclassical 
transport theory within the relaxation-time approximation, assuming 
temperature-dependent but field-independent mean free paths ($\lambda$)
and $\Delta E_F$ = 4 meV. At 50~K, we assume isotropic $\lambda$'s 
for holes, which are 40 nm for the $\alpha$ and $\alpha'$ pockets
and 10 nm for the other smaller pockets, and anisotropic $\lambda$ 
for electrons ($\beta$), which are 20 and 80 nm for motion along 
and perpendicular to the CuO chain, respectively. At 1.5~K they are 
increased to ten times the values at 50~K. These values of 
$\lambda$'s are adjusted to show a theoretical reproduction of the 
experimental data although they are quite a bit larger than 
those estimated from oscillation amplitudes. With the assumed $\lambda$'s,
the calculated Hall coefficient is negative at high magnetic field and 
low temperature (Fig.~3a), becoming positive at high temperature 
(Fig.~3b), as in the experiment\cite{leboeuf}.

The SdH oscillations are displayed in Fig.~3a by modifying the 
conductivity tensor $\sigma_{ij}$ to include effects of the Landau 
levels\cite{tando}. The above mentioned mean free paths ($\lambda$) 
are used for $\sigma_{ij}$ itself; however, a reduction of $\lambda$'s 
by a factor of 0.05 is assumed for the modification factor 
of $\sigma_{ij}$ for quantum oscillations, yielding SdH oscillation 
amplitudes close to experiments\cite{leboeuf}. This may suggest that the 
charge carriers in the material lose their quantum coherence much faster 
than their classical linear momenta, but it is beyond the scope of 
our present work to justify the assumed $\lambda$'s. In our results, 
the low-frequency oscillations (from the hole pockets) are much stronger 
than the high-frequency oscillation (from the electron pocket) 
since the average mean free {\em time} is larger for the holes than 
for the electrons even with the assumed $\lambda$'s because of the 
difference in their group velocities. Thus, the dominant oscillation 
in the Hall coefficient (Fig.~3a) originates from the hole pockets 
although the Hall coefficient itself is negative at high field due 
to the electron pocket. The calculated SdH oscillations grow with the 
magnetic field (Fig.~3a), as observed experimentally\cite{leboeuf}.

Compared with model calculations considering magnetic fluctuations
\cite{harrison,morinari}, our results show that the presence of CuO 
chains in YBa$_2$Cu$_3$O$_{6.5}$ is important for explaining the 
high-frequency quantum oscillation and the negative Hall coefficient. 
Since our results are based on a static long-range magnetic order 
stable in the LSDA+$U$ method, fluctuations in real materials may 
modify the FS. As discussed above, one possibility is the evolution 
of the small-size hole pockets ($\alpha$ and $\alpha'$ in Fig.~1d) 
to arcs, as proposed by the model calculations.

While the $d$-density-wave theory predicts hole pockets larger 
than electron pockets, our result predicts an electron pocket larger 
than hole pockets, and assigns FS pockets to the observed frequencies 
oppositely. Thus, in our work, the observed major frequency originates
from small-size hole pockets while the negative Hall coefficient is 
due to a large-size electron pocket.


\begin{figure} 
\centering
\epsfig{file=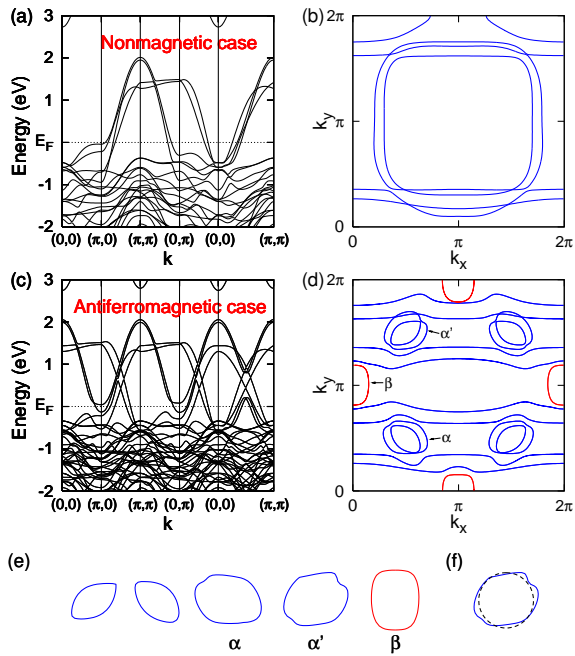,width=7.5cm,angle=0,clip=}  
\caption{(Color online) 
Electronic structures in YBa$_2$Cu$_4$O$_8$ with Coulomb interaction 
at Cu $d$ orbitals ($U$ = 3.1 eV and $J$ = 0.8 eV). (a) The LDA+$U$ 
band structure and (b) the FS in the non-magnetic case. (c) The 
LSDA+$U$ band structure and (d) the FS with a checker-board-type 
AFM order in the CuO$_2$ planes. In (b) and (d), the FSs are drawn in 
the momentum space corresponding to a real-space unit cell 
(0.384 $\times$ 0.387 nm$^2$) containing a Cu atom on each CuO$_2$ 
plane. Blue (red) lines are hole pockets and open orbits (electron 
pockets). (e) FS pockets in (d). The $\alpha$, $\alpha'$, and 
$\beta$ pocket areas are 721, 729, and 685 T, respectively. 
(f) Comparison with the measured FS pocket area 
(660~T \cite{bangura,yelland}) in the dashed line.
}
\end{figure}

For YBa$_2$Cu$_4$O$_8$, as in the case of YBa$_2$Cu$_3$O$_{6.5}$, we 
obtain very different FSs with and without AFM order in the CuO$_2$ 
planes (Fig.~4). With the LDA+$U$ method with $U$ = 3.1 eV and 
$J$ = 0.8 eV for all Cu $d$ orbitals, we obtain a FS consisting 
of large hole pockets and open orbits (Fig.~4b) which is consistent 
with previous first-principles calculations\cite{yu}. When the 
magnetic order is considered by the LSDA+$U$ method with the same 
$U$ and $J$, AFM order is stable in the CuO$_2$ planes with 
0.22 $\mu_B$ at each Cu atom, and the FS consists of small hole 
pockets ($\alpha$ and $\alpha'$ indicating the two largest ones), 
open orbits, and small electron pockets ($\beta$), as shown in Fig.~4d. 
The calculated FS pocket areas are 721~T ($\alpha$), 729~T ($\alpha'$), 
and 685~T ($\beta$), which are close to the 
measured value 660~T \cite{bangura,yelland}, overestimating 
it by about 10 \% or less.  Contrary to YBa$_2$Cu$_3$O$_{6.5}$, 
the calculated FS pocket sizes in YBa$_2$Cu$_4$O$_8$ are 
sensitive to $U$ and $J$ around the used values. Calculated 
cyclotron effective masses, 0.45 $m_e$ for holes and 0.52 $m_e$ 
for electrons, are much smaller than the measured values of 
2.7 $\sim$ 3.0 $m_e$ \cite{bangura,yelland}, but calculated Sommerfeld 
coefficient for the normal-state specific heat, 
6.97 mJ mol$^{-1}$ K$^{-2}$, is rather close to the experimental 
value of 9 mJ mol$^{-1}$K$^{-2}$ \cite{loram2}.

In summary, we have studied the electronic structures of 
YBa$_2$Cu$_3$O$_{6.5}$ and YBa$_2$Cu$_4$O$_8$ by the LSDA+$U$ method, 
and the results yield FS topologies fully consistent with quantum 
oscillation measurements. It is shown that the magnetic order in the 
CuO$_2$ planes may explain quantitatively the dHvA and SdH oscillation 
frequencies, the negative Hall coefficients, and the specific heat. 
These results point to the importance of magnetic order for 
understanding the electronic structures of the underdoped 
high-$T_c$ cuprates.

This work was supported by the NRF of Korea 
(Grant Nos. KRF-2007-314-C00075 and R01-2007-000-20922-0), 
by NSF under Grant No. DMR07-05941, and by the Director, 
Office of Science, Office of Basic Energy Sciences, 
Materials Sciences and Engineering Division, U.S. DOE
under Contract No. DE-AC02-05CH11231. Early version of LSDA+$U$ 
methodology was supported by NSF; H.J.C.
was supported by BES DMSE during collaborative visits.
Computational resources have been provided by KISTI Supercomputing 
Center (Project No. KSC-2008-S02-0004), NSF through TeraGrid 
resources at SDSC, and DOE at LBNL's NERSC facility.

\end{document}